\documentclass[PRL,twocolumn,showpacs,preprintnumbers,nofootinbib,amsmath,amssymb,floatfix]{revtex4}
\usepackage{graphicx}
\usepackage{dcolumn}
\usepackage{bm}

\newcommand{\beq}{\begin{equation}}
\newcommand{\eeq}{\end{equation}}
\newcommand{\beqa}{\begin{eqnarray}}
\newcommand{\eeqa}{\end{eqnarray}}

\begin{document}

\title{Constraining Asymmetric Bosonic Non-interacting Dark Matter with Neutron Stars}
\author{Yi-Zhong Fan,~Rui-Zhi Yang,~and Jin Chang}
\affiliation{%
\ Purple Mountain Observatory, Chinese Academy of Sciences, Nanjing 210008, China \\
\ Key Laboratory of Dark Matter and Space Astronomy, Chinese Academy of Sciences, Nanjing 210008, China\\
}
\begin{abstract}
The Hawking evaporation of small black holes formed by the collapse of dark matter at the center of neutron stars plays a key role in loosing the constraint on the mass of asymmetric bosonic non-interacting dark matter particles.
Different from previous works we show that such a kind of dark matter is viable in the mass range from $3.3$ GeV to $\sim 10$ TeV, which covers the most attractive regions, including the preferred asymmetric dark matter mass $\sim 5.7$ GeV as well as the $5-15$ GeV range favored by DAMA and CoGeNT.
\end{abstract}
\pacs{95.35.+d, 95.30Cq, 97.60.Jd}
\maketitle

\section{Introduction}
There is compelling evidence from astrophysical and cosmological data that the dominant
component of the matter in the universe is in the form of ``dark matter" that only
interacts very weakly with ordinary matter \cite{Bertone05}. One of the remarkable features of the standard cosmological model is
that the dark matter density today is comparable with that of the baryons
(i.e., $\rho_{\rm DM}\sim 5.7\rho_{\rm baryon}$) \cite{Eisenstein05}. Such a fact could be naturally explained by
an asymmetry in dark matter similar to that in baryons \cite{Gelmini87}. Though higher or lower masses are possible
\cite{Hooper05,Falkowski11}, the asymmetric dark matter (ADM) models prefer to
have DM mass $\sim \rho_{\rm DM}m_{\rm n}/\rho_{\rm baryon}\sim 5.7$ GeV \cite{Kaplan09}, which is within the inferred mass range
($5-15$ GeV) of the dark matter candidates found by DAMA \cite{DAMA2008} and CoGeNT \cite{Aalseth11}, where $m_{\rm n}$ is the mass of neutrons.
In the ADM scenarios, despite the DM particles
can in principle have weak-scale interactions and therefore sizable scattering cross sections off
baryons, annihilations do not occur because of the presence of an asymmetry in the DM sector
between particles and anti-particles. Therefore the ADM particles can be captured by astrophysical
bodies such as the sun, earth, white dwarfs and neutron stars \cite{press85,goldman89,Bertone08}. The captured particles
will eventually be thermalized and centered in the core of the star, namely, a dense dark matter core forms \cite{press85,goldman89,Bertone08}. Due to the non-annihilation of ADM particles, the build up of a concentration at the center of neutron stars may lead up
 to a collapse of these particles into a small black hole and the neutron star may be destroyed. If the central
temperature of very old isolated neutron stars falls below the critical temperature to form a Bose-Einstein
condensate (BEC), the bosonic particles in the ground state condense and no longer follow the
thermal distribution. Consequently the process of collapse will be significantly enhanced. Therefore the existence of very old neutron stars in the Galaxy imposes a valid constraint on non-annihilating bosonic weakly interacting massive particles (WIMPs). In the initial work by Goldman and Nussinov \cite{goldman89} a mass larger than $\sim 1~{\rm GeV}$ has been excluded. In other words, the preferred ADM models with a mass $\sim 5.7$ GeV can not be non-annihilating and bosonic. Such a result is a piece of evidence that the compact objects can be valuable probe of the physical properties of dark matter particles.
Recently, after taking into account the Hawking evaporation of the newly-formed small black hole \cite{Hawking1974,MacGibbon90}, Kouvaris and Tinyakov  \cite{bosondm} have shown that the non-interacting ADM particles can be bosonic as long as they have a mass $m_{\rm \chi} \geq 16$ GeV, i.e., the allowed mass range of non-interacting bosonic ADM particles has been considerably extended. Such a progress is remarkable but the $5-15$ GeV range favored by DAMA and CoGeNT has still been excluded. In this work we show that with the advanced estimate of Hawking evaporation rate \cite{MacGibbon90} as well as the rate of accretion onto the black holes formed at the center of neutron stars,  asymmetric bosonic non-interacting dark matter with a mass $>3.3$ GeV is viable, which covers the most interesting regions,  including the preferred ADM mass $\sim 5.7$ GeV as well as the $5-15$ GeV range favored by DAMA and CoGeNT. We also find that the asymmetric bosonic non-interacting dark matter particles with a mass above $\sim 10$ TeV has been excluded by current neutron star observations.

\section{Constraint on the mass of asymmetric non-interacting bosonic dark matter}
WIMPs scatter with nuclei and then lose their kinetic energy when
pass through a star. If the kinetic energy is
smaller than the stellar gravitational potential the dark matter particles will get captured \cite{press85} and concentrate in the center within the radius
\cite{ther}
\begin{equation}
r_{\rm th}
\approx 74~{\rm cm}
T_{\rm c,6}^{1/2}(\frac{100\rm GeV}{m_{\rm \chi}})^{1/2}\rho_{\rm c,15}^{-1/2},
\end{equation}
where $T_{\rm c}$ ($\rho_{\rm c}$) is the temperature (density) of the star core. Through this work we denote $\rho_{\rm c,15}={\rho_{\rm c}/10^{15}~{\rm g~cm^{-3}}}$ and $T_{\rm c,6}=T_{\rm c}/10^{6}$ K.

The mass of the ordinary matter contained in such a volume is
\begin{equation}
m_{\rm self}\approx 10^{45}~{\rm GeV}~T_{\rm c,6}^{3\over 2}(\frac{100\rm GeV}{m_{\rm \chi}})^{3\over 2}\rho_{\rm c,15}^{-1\over 2}.
\end{equation}
The mass of the captured dark matter can be estimated as \cite{Bertone08}
\begin{equation}
m_{\rm dm}\approx 1.3 \times 10^{39}~{\rm GeV}~f(\frac{\rho_{\rm dm}}{\rm 0.3~GeV~cm^{-3}})(\frac{t}{\rm 10^{6}~yr}),
\end{equation}
where $\rho_{\rm dm}$ is the energy density of dark matter surrounding the neutron star, $f\equiv \min\{1,~\sigma_{\chi,-44.7}\}$ and $\sigma_\chi$ is the spin-independent cross section of the  dark matter scattering with the neutron/proton.

\subsection{Excluding the light asymmetric bosonic dark matter}
For bosonic WIMPs, a Bose-Einstein condensate (BEC) may be formed. Since such a state is more compact, the self-gravitation regime starts much earlier. The particle density required to form BEC is
$n_{\rm BEC}\simeq 1.5\times 10^{33}~{\rm cm^{-3}} (\frac{m_{\rm \chi}}{100\rm GeV})^{3/2}({T_{\rm c}\over 10^{6}~{\rm K}})^{3/2}$
and the total mass of the needed particles is
\begin{equation}
m_{_{\rm BEC}}=2.5\times 10^{41}~{\rm GeV}~({m_\chi \over 100~{\rm GeV}})T_{\rm c,6}^{3}\rho_{\rm c,15}^{-3/2}.
\end{equation}

The size of the condensed state is determined by the radius of the wave function of the WIMP ground state in the gravitational potential of the star, i.e.,
\begin{equation}
r_{\rm cond}\approx 1.6\times 10^{-5}~{\rm cm}~({{100~\rm GeV} \over m_\chi} )^{1/2},
\end{equation}
which is much smaller than $r_{\rm th}$. The self-gravitation regime reaches for a mass of the condensed matter
\[
m_{\rm BEC,self}\approx 8\times 10^{24}~{\rm GeV}~(m_\chi/100~{\rm GeV})^{-3/2}\rho_{\rm c,15}.\]

For non-interacting bosons 
only the uncertainty principle opposes the collapse. As long as the mass of self-gravitating bosons reaches \cite{bosondm}
\begin{equation}
m_{\rm c}\approx {2\over \pi}{m_{\rm pl}^{2}\over m_{\rm \chi}}=9.5\times 10^{35}~{\rm GeV}~({m_\chi \over 100~{\rm GeV}})^{-1},
\label{eq:M_c}
\end{equation}
they will collapse into a black hole,
where $m_{\rm pl}=1.22\times 10^{19}$ GeV is the Planck mass.
For $m_\chi>1$ keV that is of our interest, $m_{\rm c}\gg m_{\rm BEC,self}$. Hence it is $m_{\rm c}$ rather than $m_{\rm BEC,self}$ that
provides a key constraint on the mass of bosonic non-interacting ADM.

For a black hole to form, the condition either $m_{\rm c}<m_{\rm BEC}<m_{\rm dm}$ or $m_{\rm self}\leq m_{\rm dm}$ should be satisfied. The request $m_{\rm BEC}<m_{\rm dm}$ suggests that the neutron star should have an age $t_{\rm NS}\geq 2\times 10^{8}~{\rm yrs}~({m_\chi \over 100~{\rm GeV}})T_{\rm c,6}^{3}\rho_{\rm c,15}^{-3/2}f^{-1}({\rho_{\rm dm}\over {\rm 0.3~GeV~cm^{-3}}})^{-1}$. While $m_{\rm c}<m_{\rm BEC}$ requires
\begin{equation}
m_\chi>m_{\rm low}=0.2~{\rm GeV}~T_{\rm c,6}^{-3/2}\rho_{\rm c,15}^{3/4},
\label{eq:m_low}
\end{equation}
and then $t_{\rm NS}> 4\times 10^{5}~{\rm yrs}~T_{\rm c,6}^{3/2}\rho_{\rm c,15}^{-3/4}f^{-1}({\rho_{\rm dm}\over {\rm 0.3~GeV~cm^{-3}}})^{-1}$. Current observations of nearby highly magnetized neutron stars suggest that these objects usually cool down to a surface temperature $T_{\rm s}\sim 10^{6}$ K (correspondingly $T_{\rm c}\sim 10^{8}$ K if the relation $T_{\rm c}\approx 10^{8}~{\rm K}~[(T_{\rm s}/10^{6})^{4}/(g/2\times 10^{14}~\rm cm~s^{-2})]^{0.455}$ \cite{Gudmundsson82} still holds, where $g$ is the surface gravity of the neutron star) in a timescale $t\sim 10^{6}$ yrs \cite{Aguilera08}. If such objects can be identified in regions with $\rho_{\rm dm}>120f^{-1}~{\rm GeV~cm^{-3}}$, one has $m_{\rm low}=0.2$ MeV.

For ADM particles lighter than $\sim 0.2 ~{\rm GeV}$, it is rather hard to have $m_{\rm self}\leq m_{\rm dm}$. For example even with very extreme parameters $\rho_{\rm dm}\sim {\rm 3~TeV~cm^{-3}}$, $t\sim 10^{10}$ yrs and $T_{\rm c}\sim 10^{4}$ K, we have $m_{\rm dm}\approx 10^{47}~{\rm GeV}$, which is still smaller than $m_{\rm self}\approx 10^{48}~{\rm GeV}~(T_{\rm c}/10^{4}~{\rm K})^{3/2}(0.1~{\rm GeV}/m_\chi)^{3/2}$. Therefore current neutron star observations are {\it not} against the keV to sub-GeV non-interacting bosonic ADM and the future observations of highly magnetized neutron star in dense dark matter region may exclude the mass range from sub-MeV to sub-GeV.

After its formation, whether the black hole can survive or not depends on the competition between Hawking evaporation and simultaneous accretion of the surrounding material. Let's discuss these two processes in some detail.

{\bf The Hawking radiation} mimics thermal emission from
a blackbody with a finite size and a temperature of \cite{Hawking1974}
\begin{equation}
T_{\rm H} = 1.06~{\rm GeV}~({M_{\rm BH}\over 10^{13}~{\rm g}})^{-1},
\end{equation}
where $M_{\rm BH}$ is the mass of the black hole.
The mass-loss rate of an evaporating black hole can be expressed as
\begin{equation}
\frac{dM_{\rm BH}}{dt} = -5.34\times 10^{25}~{\rm g~s^{-1}}~F(M_{\rm BH})({M_{\rm BH}\over 1~{\rm g}})^{-2},
\end{equation}
where $F(M_{\rm BH})$ accounts for degrees of freedom of each species of radiated particles and can be estimated with eq.(9) of \cite{MacGibbon90}. When the temperature of the black hole exceeds the quark-gluon deconfinement temperature ($\sim 200$ MeV), the quark, gluon and lighter particles will be emitted. Hence the mass loss rate can be considerably larger than the
estimate $dM_{\rm BH}/{dt}=-\hbar c^{4}/15360\pi G^{2}M_{\rm BH}^{2}$ adopted in both \cite{bosondm} and \cite{boson-2}, where $G$ is Newton's constant, $c$ is the speed of light and $\hbar$ is the reduced Planck's constant.

{\bf Bondi accretion onto black hole} has been widely discussed since 1952 \cite{Bondi52}. For current purpose we need to take the relativistic approach.
Following Appendix G of \cite{shapiro87}, the rate of accretion of rest mass onto the black hole is
$4\pi m_{\rm n} nu r^{2}=\dot{M}={\rm constant}$ (independent of the radius $r$ to the black hole),
where $n$ is the baryon number density and $u$ is the inward radial velocity component.
To calculate an explicit value for $\dot{M}$, we need to adopt an equation of state and take it to be $P=K n^{\Gamma}$, where $P$ is the pressure. In \cite{Bondi52}, $K$ and $\Gamma$ are assumed to be constant. Currently the density is so high that the equation of state likely changes when $n$ reaches a critical value. For $n< n_{\rm t}=1.15~{\rm baryon~fm^{-3}}$ ($\rho_{\rm t}\approx 1.9\times 10^{15}~{\rm g~cm^{-3}}$), motivated by the latest astrophysical measurement on the equation of state of the ultra-dense material in neutron stars \cite{Ozel2010} we take $P\approx 2~{\rm MeV~fm^{-3}}~(n/0.2~{\rm baryon~fm^{-3}})^{3}$. At $n=n_{\rm t}$, $P\approx n m_{\rm n}c^{2}/3$ and quark matter likely forms (the temperature is denoted as $T_{\rm t}$), for which $\Gamma\approx 4/3$ and the sound speed is $a_{\rm s}\approx c/\sqrt{3}$. The energy density (baryon number density) is proportional to $T_{\rm t}^{4}$ ($T_{\rm t}^{3}$).
Adopting eq.(G.17) of \cite{shapiro87}, at the critical radius $r_{\rm s}\approx 3GM_{\rm BH}/c^{2}$, we have $u_{\rm s}=c/\sqrt{6}$, and $u_{\rm s}^{2}/a_{\rm s}^{2}=1/2$. Hence the relativistic Bernoulli equation (i.e., eq.(G.22) of \cite{shapiro87}) gives $T_{\rm s}/T_{\rm t}=(a_{\rm s}^{2}/u_{\rm s}^{2})^{1/2}=\sqrt{2}$, $n_{\rm s}=(a_{\rm s}^{2}/u_{\rm s}^{2})^{3/2}n_{\rm t}=2\sqrt{2}n_{\rm t}$, and then
\begin{equation}
\dot{M}=4\pi m_{\rm n} n_{\rm s}u_{\rm s} r_{\rm s}^{2}\approx 24\sqrt{3}\pi \rho_{\rm t}G^{2}M_{\rm BH}^{2}/c^{3}.
\end{equation}
Our accretion rate is lower than the preliminary estimate with classical Bondi accretion formula \cite{bosondm} by a factor of $\sim 10$.
\\

The black hole will disappear as long as $\dot{M}+\frac{dM_{\rm BH}}{dt}<0$, such a request defines a critical mass of the black hole, i.e.,
\begin{equation}
M_{\rm BH,c}=5.2\times 10^{13}~{\rm g}~({\rho_{\rm t}\over 2\times 10^{15}~{\rm g~cm^{-3}}})^{-1/4}({F(M_{\rm BH})\over 6})^{1/4}.
\end{equation}
The Hawking temperature is $T_{\rm H}\approx 203$ MeV, for which $F(M_{\rm BH})\approx 6$.

The mass of the dark matter particles should satisfy (i.e., eq.(\ref{eq:M_c}))
\begin{eqnarray}
m_\chi &>& m_{\rm \chi,c}=9.5\times 10^{37}~{\rm GeV}^{2}/ (M_{\rm BH,c}c^{2})\nonumber\\
&=& 3.3~{\rm GeV}~({\rho_{\rm t}\over 2\times 10^{15}~{\rm g~cm^{-3}}})^{1/4}({F(M_{\rm BH})\over 6})^{-1/4},
\label{eq:main-constraint}
\end{eqnarray}
otherwise the black hole will be bigger and bigger and finally the whole neutron star is destroyed.

\subsection{Excluding the TeV asymmetric bosonic dark matter}
Very old neutron stars will be destroyed too if BEC does not form (i.e., $M_{\rm dm}<M_{_{\rm BEC}}$) but the self-gravitating dark matter core is massive enough for black hole to form (i.e., $M_{\rm dm}>M_{_{\rm self}}$ and $M_{\rm dm}>2m_{\rm pl}^{2}/\pi m_{\chi}$). The existence of old neutron stars in the Galaxy thus imposes a tight constraint on the heavy ADM model \cite{boson-2}. In this case the request $M_{\rm dm}>2m_{\rm pl}^{2}/\pi m_{\chi}$  usually does not provide us new information and we focus on the first two requests, i.e.,
\begin{eqnarray}
m_\chi>m_1&=&0.5~{\rm GeV}~f{\rho_{\rm dm}\over 0.3~{\rm GeV~cm^{-3}}}{t\over 10^{6}~{\rm yr}}\nonumber\\
&&T_{\rm c,6}^{-3}\rho_{\rm c,15}^{3/2},
\end{eqnarray}
and
\begin{eqnarray}
m_\chi>m_2 &=& 840~{\rm TeV}~T_{\rm c,6}f^{-2/3}({\rho_{\rm dm}\over 0.3~{\rm GeV~cm^{-3}}})^{-2/3}\nonumber\\
&&({t\over 10^{6}~{\rm yr}})^{-2/3}\rho_{\rm c,15}^{-1/3}.
\end{eqnarray}
Therefore we need $m_\chi>m_{\rm up}\equiv \max\{m_1,~m_2\}$. Due to the relatively fixed $\rho_{\rm c}$ and the reverse sign of the indices of $m_1$ and $m_2$ on $f$, $\rho_{\rm dm}$, $t$, it is straightforward to show that $m_{\rm up}$ reaches the minimum when $m_1=m_2$, for which we have
\begin{eqnarray}
T_{\rm c,crit} &=& 2.8\times 10^{4}~{\rm K}~f^{5/12}({\rho_{\rm dm}\over 0.3~{\rm GeV~cm^{-3}}})^{5/12}\nonumber\\
&&({t\over 10^{6}{\rm yr}})^{5/12}\rho_{\rm c,15}^{11/24},
\label{eq:T_c}
\end{eqnarray}
and then
\begin{equation}
m_{\rm up}=24~{\rm TeV}~f^{-{1\over 4}}({\rho_{\rm dm}\over 0.3~{\rm GeV~cm^{-3}}})^{-{1\over 4}}({t\over 10^{6}{\rm yr}})^{-{1\over 4}}\rho_{\rm c,15}^{1/8}.
\end{equation}
Since $f^{-1/4}\geq 1$, for the extreme parameters $\rho_{\rm dm} \sim 3~{\rm TeV~cm^{-3}}$ and $t\sim 10^{10}$ yrs, the above constraint yields that $m_{\rm up}\geq 240$ GeV. However, such a tight constraint is not reachable because the required $T_{\rm c}$ is high up to $\sim 10^{8}$ K, which is impossible for a neutron star as old as the Universe. Nevertheless, even in such an extreme scenario, the resulting $m_{\rm up}\sim 240$ GeV is far larger than $m_{\rm \chi,c}\approx 3.3$ GeV, demonstrating that {\it the window for non-interacting bosonic asymmetric dark matter models does open.}

For isolated neutron stars, with the relation between $T_{\rm c}$ and $T_{\rm s}$, we have
\begin{eqnarray}
T_{\rm s,crit} &\approx & 1.1\times 10^{4}~{\rm K}~f^{0.23}({\rho_{\rm dm}\over 0.3~{\rm GeV~cm^{-3}}})^{0.23}\nonumber\\
&&({t\over 10^{6}{\rm yr}})^{0.23}\rho_{\rm c,15}^{0.25}g_{14.3}^{1/4}.
\label{eq:T_s}
\end{eqnarray}
To form a black hole, $m_{\chi}$ should satisfy $m_\chi\geq m_{\rm up}(T_{\rm s}/T_{\rm s,crit})^{-5.46}$ for $T_{\rm s}<T_{\rm s,crit}$ and $m_\chi\geq m_{\rm up}(T_{\rm s}/T_{\rm s,crit})^{1.82}$ for $T_{\rm s}\geq T_{\rm s,crit}$.
Among the parameters involved in the above equation, $T_{\rm s}$, $\rho_{\rm dm}$ and $t$ are measurable, and $\rho_{\rm c}$ and $g$ are relatively well determined. So
 the tightest constraint on $m_\chi$ is achievable.

The cooling of neutron stars is still to be better understood. The fit to the surface temperature of the best studied isolated neutron stars suggests that for a mass $\sim 1.4M_\odot$,  one likely has $T_{\rm s}\sim 5\times 10^{5}$ K ($\sim 2.3\times 10^{4}$ K) at $t\sim 10^{6}$ yrs ($\sim 2.5\times 10^{7}$ yrs) \cite{Yakovlev08}. With eq.(\ref{eq:T_s}) we have $T_{\rm s,crit} \approx 2.3\times 10^{4}~{\rm K}\sim T_{\rm s}$ at a time $t\sim 2.5\times 10^{7}$ yrs.
The existence of isolated neutron stars older than $2.5\times 10^{7}$ yrs thus requires that
\begin{equation}
m_\chi<m_{\rm up} \approx 10~{\rm TeV}~f^{-{1\over 4}}({\rho_{\rm dm}\over 0.3~{\rm GeV~cm^{-3}}})^{-{1\over 4}},
\label{eq:Constraint-TeV}
\end{equation}
otherwise such stars will have been destroyed by the black hole formed at their center (the formed black hole has a mass $\gg 5.6\times 10^{38}$ GeV, for which the mass loss due to Hawking evaporation is too small to balance the Bondi accretion. It is straightforward to show this with the formulae presented in Section II A).

\begin{figure}
\includegraphics[width=96mm,angle=0]{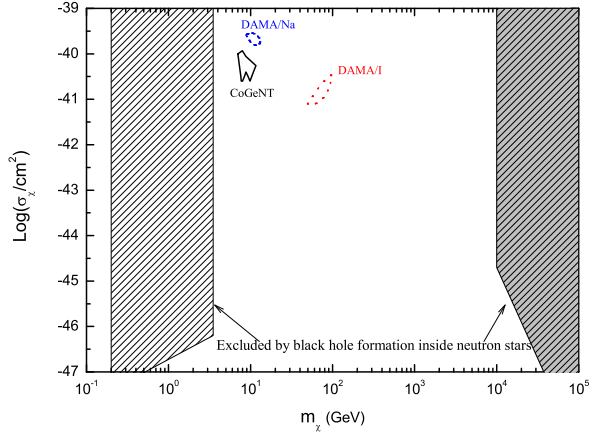}
  \caption{The exclusion regions as a function of $m_\chi$ and $\sigma_\chi$ for isolated neutron stars at local DM density $\rho_{\rm dm}=0.3~{\rm GeV~cm^{-3}}$. The left exclusion region is the case of J0108-1431 ($t_{\rm NS}\sim 1.6\times 10^{8}$ yrs and $T_{\rm c}\sim 10^{6}$ K  \cite{Mignani08}) while the right exclusion region is based on eq.(\ref{eq:Constraint-TeV}).}
  \label{fig:1}
\end{figure}

\section{Conclusion}
For the bosonic non-interacting asymmetric dark matter particles, the build up of a concentration at the center of neutron stars may lead up
 to a collapse of these particles into a small black hole and the neutron star may be destroyed. Therefore the existence of old neutron stars in the Galaxy imposes a tight constraint on such a kind of dark matter, as firstly noticed in 1989 \cite{goldman89}. A robust constraint is however not achievable until the Hawking evaporation of the newly-formed small black hole and the simultaneous accretion have been properly addressed. We find out that in previous works the Hawking evaporation rate has been underestimated while the Bondi accretion rate has been overestimated (see section II A for the details). After the corrections, the asymmetric bosonic non-interacting dark matter with a mass $>m_{\rm \chi,c}\approx 3.3$ GeV is found to be viable. We also show that for $m_\chi>m_{\rm up}\sim 10(\rho_{\rm dm}/0.3~{\rm GeV~cm^{-3}})$ TeV, the black hole can form (without reaching the Bose-Einstein condensation regime) and swallow the neutron star. Interestingly, even in the most-extreme/unrealistic scenario, we have $m_{\rm up}\sim 240$ GeV, which is much larger than $m_{\rm \chi,c}\approx 3.3$ GeV, demonstrating that there is indeed a nice window for non-interacting bosonic asymmetric dark matter models (see section II B for the details). We conclude that the asymmetric non-annihilating bosonic dark matter is viable if its mass is in the range from $3.3$ GeV to $\sim 10$ TeV, which covers the most attractive regions, including the preferred asymmetric dark matter mass $\sim 5.7$ GeV as well as the $5-15$ GeV range favored by DAMA and CoGeNT (see Fig.\ref{fig:1}). Moreover, we find out that current neutron star observations are {\it not} against the keV to sub-GeV non-interacting bosonic ADM and the future observations of magnetized neutron star in dense dark matter region may exclude the mass range from sub-MeV to sub-GeV (see eq.(\ref{eq:m_low}) and the subsequent two paragraphs for the discussion).

\acknowledgments  This work is supported in part by National Basic Research Program of China under grants 2009CB824800 and 2010CB0032, and by National Natural Science of China under grants 10920101070, 10925315 and 11073057. YZF is also supported by the 100 Talents program of Chinese Academy of Sciences.\\

Electric addresses: yzfan@pmo.ac.cn, bixian85@pmo.ac.cn, chang@pmo.ac.cn

\end{document}